\documentstyle[aps,12pt,preprint,epsfig]{revtex}
\tightenlines

\def \to {\rightarrow}
\def \beq {\begin{equation}}
\def \eeq {\end{equation}}
\def \ba {\begin{eqnarray}}
\def \ea {\end{eqnarray}}
\def \jpsi {J/\psi}
\def \< {\langle}
\def \> {\rangle}
\draft
\begin{document}

\title{Inelastic $J/\psi$ production in polarized photon-hadron collisions}
\author{Feng Yuan, Hui-Shi Dong and Li-Kun Hao }
\address{\small {\it Department of Physics, Peking University, Beijing
100871, People's Republic of China}}
\author{Kuang-Ta Chao}
\address{\small {\it China Center of Advanced Science and Technology
(World Laboratory), Beijing 100080, People's Republic of China\\
and Department of Physics, Peking University, Beijing 100871,
People's Republic of China}}
\maketitle

\begin{abstract}
Presented here is a calculation of inelastic $J/\psi$ production in
polarized photon-hadron collisions under 
the framework of NRQCD factorization formalism.
We consider the photoproduction of $\jpsi$ in the energy range relevant to
HERA.
The Weizs\"acker-Williams approximation is adopted in the evaluation of the
cross sections for $ep$ collisions.
We found that this process can give another independent test for the
color-octet mechanism, and 
the different features for the 
two color-octet processes may provide further
informations on the mechanism for inelastic $\jpsi$
photoproduction.
And the discrepancy on the production asymmetry $A$
between various sets of polarized gluon distribution functions is
also found to be distinctive.
\end{abstract}
\pacs{PACS number(s): 12.40.Nn, 13.85.Ni, 14.40.Gx}

\section{Introduction}

The nucleon spin structure has attracted a lot of attention in the
last ten years since the discovery of the so-called ``spin crisis" by the
European Muon Collaboration (EMC)\cite{emc}. 
During these times, much has been done to understand
the spin structure of proton both on the theoretical side and 
the experimental side (mostly from the polarized DIS data). 
However, the use of DIS data alone does not allow an accurate 
determination for the polarized parton densities, especially for the 
polarized gluon distribution, which is crucial important to
understand the gluon contribution to the proton spin.
And now, the upgrade of HERA collider with polarized 
electron beam and proton beam at $\sqrt{s}=300GeV$
in the near future has been considered for a long time\cite{phera}.
With this high energy polarized electron-proton collider, one can 
investigate the polarized structure functions in much broader range of
$x$ and $Q^2$, especially, to very low $x$ and high $Q^2$ region.
And also the exploration of
polarized gluon distribution $\Delta G(x,Q^2)$ will become accessible.
To determine $\Delta G(x,Q^2)$, various photon-gluon fusion processes 
can be used.
The typical promising process for this purpose
is the large$-p_T$ di-jet production\cite{phera}, which is a gluon induced process
even in the leading order (LO) of QCD calculations.
However, for this process one must deal with the uncertainty arising from
the resolved photon contributions.
In this paper, we will consider the inelastic $\jpsi$ production in the
polarized photon-hadron collisions.
Because the inelastic $\jpsi$ photoproduction is also a gluon induced
process, so this process can be viewed as a possible way to 
study the polarized gluon distribution in the proton.
Furthermore, the resolved photon contribution to this process is negligible
in the major kinematical region of the produced $\jpsi$.
So, by inducing some kinematical cut, the uncertainties due to the resolved 
photon contribution can be eliminated safely.

On the other hand, studies of heavy quarkonium production in high energy 
energies is another hot topic in recent years.
In the conventional picture, the heavy quarkonium production is described in 
the color-singlet model\cite{cs}.
In this model, it is assumed that the heavy quark pair must be 
produced in a color-singlet state in short distance with the same 
angular-momentum quantum number as the charmonium meson which is 
eventually observed.
However, with the recent Tevatron data on high $p_T$ $\jpsi$ production
appearing, this color-singlet picture for heavy quarkonium production 
became questionable,
The data is larger than the theoretical prediction of the color-singlet
model by a factor of about $30\sim 50$\cite{fa}. This is called the 
$\jpsi$ ($\psi'$)  surplus problem.
On the theoretical side, the naive color-singlet model
has been supplanted by the nonrelativistic QCD (NRQCD) factorization 
formalism \cite{GTB95}, which allows the infrared safe calculation
of inclusive charmonium production and decay rates.
In this approach, the production process is
factorized into short and long distance parts, while the latter is
associated with the nonperturbative matrix elements of four-fermion
operators. 
So, for heavy quarkonium production, the quark-antiquark pair does not 
need to be in the color-singlet state in the short distance production stage,
which is at the scale of $1/m_Q$ ($m_Q$ is the heavy quark mass).
At this stage, it allows the other color configuration than the singlet 
for the heavy quark pair, such as color-octet.
The later situation for heavy quarkonium production is called the 
color-octet mechanism.
In this production mechanism, heavy quark-antiquark pair is produced
at short distances in a color-octet state, and then hadronizes into a 
final state quarkonium (physical state) nonperturbatively. 
Provided this color-octet mechanism, one might provide explain the
Tevatron data on the surplus of $J/\psi $ and $\psi ^{\prime }$ production\cite{surplus,s1}.

Even though the color-octet mechanism has achieved some successes in 
describing the production and decay of heavy quarkonia, more tests of
this mechanism are still needed.
Recently, the photoproduction data from HERA \cite{photo,zeus} puts
a question about the universality of the color-octet matrix elements
\cite{photo2,MK}, in which the fitted values of the matrix elements 
$\langle {\cal O}^{J/\psi}_8 ({}^1S_0)\rangle$ and $\langle {\cal
O}^{J/\psi}_8({}^3P_J)\rangle$ are one order of magnitude smaller than
those determined from the Tevatron data \cite{s1}.
(More recently, possible solutions for this problem have been suggested 
in\cite{explain,explain2}.)
In this situation it is certainly helpful to find
other processes to test the color-octet mechanism in the heavy 
quarkonium production. 
So, the inelastic $\jpsi$ production in polarized photon-hadron collisions
can give another independent test of the production mechanism, aside
from the unpolarized $\jpsi$.

The rest of the paper
is organized as follows. In Sec.~II, we will give the polarized 
cross section formula for the inelastic $\jpsi$ production at the polarized 
electron-proton collider. Here, we adopt the Weizs\"acker-Williams
approximation to calculate the electroproduction cross section with the
photoproduction cross section. The numerical results are given in Sec.~III.
We will display the polarized cross sections for both the polarized 
photon-proton collisions and the electron-proton collisions.
All of our results are relevant to the energy range of the upgrade 
polarized HERA collider. 
We also study the $\jpsi$ production asymmetry and its dependence on the
polarized gluon distributions.
In Sec.~IV, we give the conclusions.

\section{polarized cross section formulas}

In photoproduction processes, the electron-proton cross section can be 
equivalent to the photoproduction cross section convoluted by the 
photon flux factor from the electron under the Weizs\"acker-Williams
approximation\cite{ww},
\begin{equation}
d\sigma_{ep}=\int dy f_\gamma^{(e)}(y)d\sigma_{\gamma p}.
\end{equation}
Here, $y$ is the energy fraction of the electron carried by the photon.
$f_\gamma^{(e)}$ is the flux factor of photon (or the distribution 
function of photon in the electron). In the following calculations,
we use the form from \cite{wwnew},
\begin{equation}
\label{fr}
f_\gamma^{(e)}(y)=\frac{\alpha_{em}}{2\pi}[\frac{1+(1-y)^2}{y}{\rm log}
	\frac{Q_{max}^2(1-y)}{m_e^2y^2}+2m_e^2y(\frac{1}{Q_{max}^2}-
	\frac{1-y}{m_e^2y^2})],
\end{equation}
where $m_e$ is the electron mass, and $Q_{max}^2$ is the maximum value
of $Q^2$ for the photoproduction processes. In the following numerical
calculations, we will adopt $Q_{max}^2=1GeV^2$.
Compared with the original Weizs\"acker-Williams function, the above 
equation contains a non-logarithmic term with a singular behavior at
$y\to 0$\cite{wwnew}. This term will give non-negligible contributions to 
the unpolarized cross section.

Correspondingly, the polarized electroproduction cross section has the
similar form,
\begin{equation}
d\Delta\sigma_{ep}=\int dy \Delta f_\gamma^{(e)}(y)d\Delta\sigma_{\gamma p},
\end{equation}
where the polarized photon flux factor function $\Delta f_\gamma^{(e)}(y)$
has been derived in Ref.~\cite{pdijet},
\begin{equation}
\label{dfr}
\Delta f_\gamma^{(e)}(y)=\frac{\alpha_{em}}{2\pi}[\frac{1-(1-y)^2}{y}{\rm log}
	\frac{Q_{max}^2(1-y)}{m_e^2y^2}+2m_e^2y^2(\frac{1}{Q_{max}^2}-
	\frac{1-y}{m_e^2y^2})].
\end{equation}
This function also contains a non-logarithmic term, which however is not 
singular at $y\to 0$. The numerical calculations show that this term will
contribute about $5\%$ to the total cross section of the inelastic $\jpsi$ 
photoproduction at $\sqrt{s}=300GeV$ for $Q_{max}^2=1GeV^2$.
The polarized cross section $\Delta \sigma$ in the above equation 
is defined in terms of $\sigma(\lambda_1,\lambda_2)$ with definite 
helicities for incoming particles,
\begin{equation}
\label{dsig}
d\Delta \sigma=\frac{1}{4}[d\sigma(+,+)+d\sigma(-,-)-d\sigma(+,-)-
	d\sigma(-,+)].
\end{equation}

In our study, we also adopt the Weizs\"acker-Williams approximation
described above to evaluate the inelastic photoproduction of 
$\jpsi$  in electron-proton collisions.
That is, the polarized and unpolarized cross sections for 
$\jpsi$ production in electron-proton collisions have the following 
forms,
\begin{eqnarray}
\label{xs}
d\Delta\sigma(ep\to e+\jpsi +X)&=&\int_{y_{min}}^{y_{max}}dy
	 \Delta f_\gamma^{(e)}(y)d\Delta\sigma(\gamma p\to \jpsi +X),\\
d\sigma(ep\to e+\jpsi +X)&=&\int_{y_{min}}^{y_{max}}dy
	 f_\gamma^{(e)}(y)d\sigma(\gamma p\to \jpsi +X),
\label{dxs}
\end{eqnarray}
where $y_{min}$ and $y_{max}$ are the kinematical boundary for the photon
energy fraction from the electron beam.
In the above equation, we do not include the resolved photon contribution
to the $\jpsi$ photoproduction at $ep$ colliders. This is because the 
resolved contribution is important only in the lower $z$ region 
($z<0.2$)\cite{pjpsi}, where $z$ is the energy fraction of the produced $\jpsi$.
So, by inducing a lower bound cut on $z$, the resolved photon contribution 
can be eliminated.
$\Delta\sigma(\gamma p\to \jpsi +X)$ is the polarized cross section for the
inelastic production of $\jpsi$ in the polarized photon-proton fusion
processes.

At high energies, the photoproduction processes proceed predominantly 
through photon-gluon fusion. So, the polarized cross section 
$\Delta\sigma(\gamma p\to \jpsi +X)$ is related to the polarized gluon 
distributions in the proton,
\begin{equation}
d\Delta \sigma(\gamma p\to \jpsi +X)=\int dx \Delta G(x,Q^2)
	d\Delta \sigma(\gamma g\to \jpsi +X).
\end{equation}
The polarized gluon distribution function $\Delta G(x,Q^2)$ is defined
as usual way, and now there are various parametrizations for this quantity,
such the GRVS sets\cite{grvs}, the GS sets\cite{gs}, the DSS sets\cite{dss}.

In photon-gluon fusion processes, according to the NRQCD factorization
formalism, the
unpolarized cross section and the polarized cross section can be
factorized as the following forms,
\begin{eqnarray}
d\sigma(\gamma+ g\rightarrow \jpsi+X)&=&\sum\limits_n F(\gamma+g\rightarrow n+X)\langle {\cal
O}_n^{\jpsi}\rangle  ,\\
d\Delta\sigma(\gamma+ g\rightarrow \jpsi+X)&=&\sum\limits_n \Delta F(\gamma+g\rightarrow n+X)\langle {\cal
O}_n^{\jpsi}\rangle.
\end{eqnarray}
Here, $n$ denotes the $c\bar c$ pair configuration in the intermediate 
state (including angular momentum $^{2S+1}L_J$ and color index 1 or 8). $F(\gamma+g\rightarrow n+X)$ and $\Delta F(\gamma+g\rightarrow n+X)$
are the short distance coefficients for the unpolarized and 
polarized cases respectively for the subprocess
$\gamma +g\rightarrow n+X$. $\langle {\cal O}_n^{\jpsi}\rangle $ is the long distance
non-perturbative matrix element which represents the probability of the
$c \bar c$ pair in $n$ configuration evolving into the physical state
$\jpsi$. The short distance coefficients $F$ and $\Delta F$ can be calculated by using
perturbative QCD in expansion of powers of $\alpha_s$. The long distance
matrix elements are still not available from the first principles at
present. However, the relative importance of the contributions from
different terms can be estimated by using the NRQCD velocity
scaling rules.

In order to suppress the diffractive contribution and higher-twist
corrections, we induced the lower bound cut for the $\jpsi$ transverse 
momentum $p_T>1GeV$. With this constraints (non-zero $p_T$
for $\jpsi$), there are the following $2\to 2$ partonic
processes contributing to $\jpsi$ inelastic photoproduction in the 
photon-gluon fusion processes,
\begin{eqnarray}
\label{cs}
\gamma +g&\rightarrow& c\bar{c}({}^3S_1,\b{1})+g,\\
\label{1s0}
\gamma +g&\rightarrow& c\bar{c}({}^1S_0,\b{8})+g,\\
\gamma +g&\rightarrow& c\bar{c}({}^3S_1,\b{8})+g,\\
\gamma +g&\rightarrow& c\bar{c}({}^3P_J,\b{8})+g.
\label{3pj}
\end{eqnarray}
The first subprocess is the color-singlet process, and the last three 
processes are all the color-octet processes.
The unpolarized cross sections for these processes have been calculated 
in literature\cite{cs,photo2,pjpsi}. The polarized cross section for the 
color-singlet process has also been calculated in \cite{cs-p}.

To calculate the polarized cross sections for the color-octet processes,
we employ the {\it helicity amplitude} method.
With this method, one can calculate the cross sections $\sigma(\lambda_1,\lambda_2)$
with definite helicities for the incident particles.
Following \cite{ham}, we choose the polarization vectors for the incident 
photon, the incident gluon and the outgoing gluon as the following forms,
\begin{eqnarray}
\label{geprc}
\label{e1}
\not\! e_1^{(\pm)}=N_e[\not\! p_1\not\! p_2\not\! p_3(1\mp\gamma_5)+
        \not\! p_3\not\! p_2\not\! p_1(1\pm\gamma_5)],\\
\label{e2}
\not\! e_2^{(\pm)}=N_e[\not\! p_2\not\! p_3\not\! p_1(1\mp\gamma_5)+
        \not\! p_1\not\! p_3\not\! p_2(1\pm\gamma_5)],\\
\label{e3}
\not\! e_3^{(\pm)}=N_e[\not\! p_3\not\! p_1\not\! p_2(1\mp\gamma_5)+
        \not\! p_2\not\! p_1\not\! p_3(1\pm\gamma_5)].
\end{eqnarray}
Where $p_1,~p_2,~p_3$ and $e_1,~e_2,~e_3$ are the momenta and the 
polarization vectors for the incident photon, incident gluon and outgoing 
gluon respectively.
And the normalization factor $N_e$ is
\begin{equation}
N_e=\frac{1}{\sqrt{2\hat s\hat t\hat u}},
\end{equation}
where the Mandelstam invariants $\hat s,~\hat t,~\hat u$ are defined as
\begin{equation}
\hat s=(p_1+p_2)^2,~~~~\hat t=(p_2-p_3)^2,~~~~\hat u=(p_1-p_3)^2,
\end{equation}
and they satisfy the relation
$$\hat s+\hat t+\hat u=M^2=4m_c^2.$$

With these definitions for the polarization vectors of the photon and gluons
Eqs.~(\ref{e1})-(\ref{e3}), 
the calculations of the helicity amplitudes are straightforward.
From these helicity amplitudes, we can easily get the cross sections with
the definite helicities for the incident particles. And then we can obtain 
the polarized cross sections $d\Delta\sigma(\gamma g\to \jpsi +X)$ for the 
different partonic processes Eqs.~(\ref{1s0})-(\ref{3pj}) from
the definition Eq.~(\ref{dsig}).
Their analytic expressions are listed in the Appendix. 
The unpolarized cross section for every partonic processes can also be 
obtained by summing up the cross sections with all of the possible
helicity states for the external particles.
We have checked that our results for the unpolarized cross sections
are consistent with the results of Ref.\cite{photo2,pjpsi}.
For comparison, we also list the unpolarized cross sections in the Appendix.

After getting the polarized cross section, we can also calculate the
asymmetry
\begin{equation}
\label{asy}
A(\beta)=\frac{d\Delta\sigma/d\beta}{d\sigma/d\beta},
\end{equation}
where $\beta$ represents for some kinematical parameters, such as $\jpsi$
transverse momentum $p_T$ and the energy fraction $z$. $z$ is defined as
$z=p\cdot k_{\jpsi}/p\cdot p_1$ with $p$, $k_{\jpsi}$, $p_1$ being the 
momenta of the proton, the outgoing $\jpsi$ and the incident photon 
respectively.
The production asymmetry parameter $A$ is more efficiency than the polarized cross section 
$\Delta \sigma$ for the study of the polarized processes in the polarized collisions.
This is because $A$ is normalized, and then less depends on the input parameters
which will strongly affect the normalization of the cross sections, such as the charm quark mass,
strong coupling, and the factorization and renormalization scales.

\section{Numerical results}

Provided the polarized and the unpolarized cross sections for the 
subprocesses, the inelastic $\jpsi$ production rate and asymmetry in 
polarized photon-hadron collisions can be obtained. For the numerical
evaluation, we choose the strong coupling constant $\alpha_s$, the
charm quark mass $m_c$, and the factorization scale $\mu^2$ to be
$$\alpha _s =0.3,~~~ m_c=1.5~GeV,~~~\mu^2=( 2m_c)^2 .$$
For the parton distribution function of the proton, we use the GRV LO
parametrization\cite{grv}.
For the polarized gluon distribution function, we will consider the GRVS sets
\cite{grvs}, the GS sets\cite{gs}, and the DDS sets\cite{dss}.

We first display the different partonic processes contributions to the
polarized cross section of $\jpsi$ inelastic production, where we
adopt the GRVS STD set as the default set
for the polarized gluon distribution function of the proton.
Fig.~1 is for the photon-proton collisions, where we typically set the energy
of photon-proton system to be $W_{\gamma p}=100GeV$ relevant to the
photoproduction processes at HERA.
Fig.~2 is for the electron-proton collisions,
where we set the energy range also relevant to HERA, $\sqrt{s}=300GeV$.
For the electroproduction cross sections, the integral region of $W_{\gamma p}$
is set to be: $(25GeV)^2<W^2_{\gamma p}<(180GeV)^2$ typically used in the
photoproduction processes at HERA.
In these two figures, we both plot the $z$ and $p_T^2$ distributions:
(a) are $z$ distributions and (b) are $p_T^2$ distributions.
To evaluate these curves, we have imposed a cut on $z$ and $P_T^2$:
$0.1<z$ for the $p_T^2$ distributions and $P_T^2>1GeV^2$ for the $z$
distributions. For the NRQCD long distance matrix elements, we set as
\begin{eqnarray}
\langle {\cal O}_1^{\psi}({}^3S_1)\rangle &=&1.16GeV^3,\\
\langle {\cal O}_8^{\psi}({}^3S_1)\rangle &=&1.06\times 10^{-2}GeV^3,\\
\langle {\cal O}_8^{\psi}({}^1S_0)\rangle &=&3.0\times 10^{-2}GeV^3,\\
\langle {\cal O}_8^{\psi}({}^3P_0)\rangle/m_c^2 &=&1.0\times 10^{-2}GeV^3,\\
\langle {\cal O}_8^{\psi}({}^3P_J)\rangle &=&(2J+1)\langle {\cal O}_8^{\psi}({}^3P_0)\rangle.
\label{hss}
\end{eqnarray}
The last equation comes from the heavy quark spin symmetry of NRQCD.
The color-singlet matrix element $\langle {\cal O}_1^{\psi}({}^3S_1)\rangle$
can be related to the wave function at origin, and can be taken
its value from the leptonic decay width of $\jpsi$.
The value of the color-octet matrix element  
$\langle {\cal O}_8^{\psi}({}^3S_1)\rangle $ is taken from a fit to the
large $p_T$ $\jpsi$ production at the Tevatron\cite{benek}.
This matrix element is not important to $\jpsi$ photoproduction
both for the unpolarized and polarized cases.
On the other hand, the two other color-octet matrix elements,
$\langle {\cal O}_8^{\psi}({}^1S_0)\rangle $
and $\langle {\cal O}_8^{\psi}({}^3P_0)\rangle$, 
are known to be very important in the inelastic $\jpsi$ photoproduction
\cite{photo2,pjpsi}.
However, their values are not well
determined from the present experimental data on $\jpsi$ productions.
Here, we just follow Ref.~\cite{pjpsi} and take their values tentatively as
listed above (which are also consistent with the naive NRQCD velocity scaling
rules) to see what their contributions to the polarized cross
section of the inelastic $\jpsi$ photoproduction will be.
The solid curves in these two figures are for the color-singlet process
contributions, the dashed lines for the color-octet ${}^1S_0$ contributions,
the dotted lines for the color-octet ${}^3P_J$ contributions, and the dotted-dashed
lines for the color-octet ${}^3S_1$ contributions.

From Figs.~1 and 2, we can see that in the lower $z$ region of 
the inelastic $\jpsi$ production the dominant contribution comes from the
color-singlet process. In the larger $z$ region, the dominant comes from 
the color-octet ${}^1S_0$ process, and its contribution rise rapidly with
$z$ in the whole region. This property is as the same as that for the 
unpolarized cross section\cite{photo2}.
However, for the unpolarized cross section of inelastic $\jpsi$
photoproduction, we know that this rapidly rising property at large $z$
is not consistent with the experimental measurements\cite{photo}.
It is argued in \cite{pjpsi} that in this region, the soft gluon resummation
is important, which may smear the $z$ distribution in the large $z$ region.
So, the predictions for the LO calculations may be not reliable.
Correspondingly, the predictions of the polarized cross section for the
color-octet ${}^1S_0$ contributions (the dashed lines) of Figs.~1 and 2
in the large $z$ region will also be strongly affected and modified
by the higher order corrections.
For the color-octet ${}^3P_J$ processes, on the other hand,
their contributions are always smaller than the color-octet ${}^1S_0$
contributions.
And especially, their contributions at large $z$ are similar to the
color-singlet contributions, i.e., fall down rapidly with $z$
at the end point region.
This property is quite different from their contributions to the unpolarized
cross section of $\jpsi$ photoproduction\cite{photo2,pjpsi}, which also
rise rapidly at large $z$ like the color-octet ${}^1S_0$ contributions.
The $p_T^2$ distributions of Figs.~1 and 2 are always dominated by the
color-octet ${}^1S_0$ process contributions.
And their contributions are larger than the color-singlet contributions by
almost an order of magnitude in the whole region of $p_T^2$,
which is similar to their contributions to the unpolarized cross section
\cite{photo2,pjpsi}.
However, the color-octet ${}^3P_J$ contributions to the $p_T^2$ distributions
of the polarized cross sections
again have different features compared to their contributions to the
unpolarized cross sections\cite{photo2,pjpsi}.
Their contributions are much smaller than the color-octet ${}^1S_0$ contributions
in the whole region of $p_T^2$, and even smaller than the color-singlet
contributions at low $p_T^2$.
At very low $p_T^2$ ($p_T^2\sim 1GeV^2$),
their contributions to the polarized cross section $\Delta \sigma$
even give a negative value, which is not plotted in the figures.

In Figs.~3 and 4, we display the asymmetry of Eq.~(\ref{asy}) for $\jpsi$
inelastic photoproduction 
as functions of $z$ and $p_T^2$. Fig.~3 is for the photon-proton collisions,
and Fig.~4 is for the electron-proton collisions.
The solid curves are the production asymmetries in the color-singlet model.
The other two curves represent the production asymmetries after including 
the color-octet contributions in two cases of values for the color-octet
matrix elements:
(I) for the dashed lines,
$\langle {\cal O}_8^{\psi}({}^1S_0)\rangle =3.0\times 10^{-2}GeV^3$
and $\langle {\cal O}_8^{\psi}({}^3P_0)\rangle/m_c^2 =0$;
(II) for the dotted lines,
$\langle {\cal O}_8^{\psi}({}^3P_0)\rangle/m_c^2 =1.0\times 10^{-2}GeV^3$
and $\langle {\cal O}_8^{\psi}({}^1S_0)\rangle =0$.
The three curves in each diagram have the same behaviors as each other.
For the $z$ distributions, the asymmetries for all of the curves rise with $z$, 
and then fall down after reaching their maximum values.
There is no steep rising
as in the polarized and unpolarized cross section distributions when $z$ approaches
its end point for all curves.
In the whole region of $z$, the case (I) color-octet predictions are larger
than the color-singlet predictions, and 
the case (II) color-octet predictions on the other hand are similar to
the color-singlet predictions in the lower $z$ region and
even smaller in the larger $z$ region.
For the $p_T^2$ distributions, the asymmetries for the three curves all rise
with $p_T^2$.
The case (I) predictions are similar to the color-singlet predictions.
However, the case (II) predictions are smaller than the color-singlet
predictions by almost a factor of two.
Especially, in the low $p_T^2$ ($p_T^2\sim 1-2GeV^2$) region, the
production asymmetry $A$ for case (II) is very small comparable with zero.

Concluding the results displayed in the above four figures (Figs.~1-4),
we can see that the difference between
the color-singlet and color-octet predictions on the distributions behaviors
and the absolute sizes of the polarized cross sections $\Delta\sigma$
and the production asymmetries $A$ may provide some information on 
$\jpsi$ production mechanism.
That is to say, the inelastic $\jpsi$ photoproduction in polarized collisions
will give another independent test for the color-octet mechanism.
In particular, as shown in the above analysis, the different features of
the contributions from the two different color-octet processes may provide
further important informations on the mechanism for the inelastic
photoproduction of $\jpsi$.

Comparing the results of Figs.~3 and 4, we can see that the production asymmetries for the
polarized electron-proton collisions are all smaller than those for the
polarized photon-proton collisions by about one order of magnitude.
This is mostly because the polarized flux factor 
$\Delta f_\gamma^{(e)}$ in Eq.~(\ref{dfr}) has a minus sign in the coefficient
before the logarithmic term (the dominant term)
compared with the unpolarized flux factor $f_\gamma^{(e)}$ in Eq.~(\ref{fr}),
which will result in a big
difference between them when convoluted with the photoproduction
cross section by using Eqs.~(\ref{xs}) and (\ref{dxs}).

Now, we turn to study the dependence of our results on the different
parameterizations for the polarized gluon distributions in the proton.
In Fig.~5 and 6, we plot the production asymmetry of $\jpsi$ as functions of $z$ 
and $p_T^2$ with different sets of the polarized gluon distribution 
functions\cite{grvs,gs,dss} (for all these sets, we just use the LO results of them).
We have tried all of the polarized gluon distribution parametrizations
for every sets mentioned above, but found the different parametrizations
for each set give the similar results.
However, the difference between different sets are
much larger.
For representation, we choose the polarized gluon distribution
function for each set as follows: the GRVS STD parametrization for the
GRVS stes (the solid lines), GS-C for the GS sets (the dashed lines),
and DSS1 for the DSS sets (the dotted lines).
Fig.~5 are the $z$ distributions and Fig.~6 are the $p_T^2$ distributions:
(a) are for the results only from the color-singlet contribution;
(b) are for the results after including the color-octet contributions, 
where we choose the color-octet matrix
elements of  
$\langle {\cal O}_8^{\psi}({}^1S_0)\rangle$ and 
$\langle {\cal O}_8^{\psi}({}^3P_0)\rangle$ as
$$\langle {\cal O}_8^{\psi}({}^1S_0)\rangle=\langle {\cal O}_8^{\psi}({}^3P_0)\rangle/m_c^2=0.008.$$
From these two figures, we can see that the difference of the asymmetry
$A$ between the different polarized gluon distribution functions is distinctive
both for the color-singlet and color-octet predictions.
The GRVS STD set give the production asymmetry much larger than those 
for the other two sets, and the GS-C set and the DDS1 set give similar
results for the production asymmetry.

\section{conclusions}
We have calculated in this paper the inelastic $\jpsi$ production
in polarized photon-hadron collisions. We have evaluated the polarized production
rate and the production asymmetry
at polarized electron-proton colliders relevant to the HERA
energy range. We performed our results both for the color-singlet
model predictions and the predictions after including the color-octet
contributions. 
We found that the inelastic $\jpsi$ photoproduction in polarized collisions
may give another independent test for the color-octet mechanism.
Especially, the different features of the contributions from the
two different color-octet processes may provide further important
informations on the mechanism for the inelastic
photoproduction of $\jpsi$.
And most important, we found that the discrepancy on the production asymmetry
$A$ between the various sets of the polarized gluon distribution function is 
distinctive.
So, the process studied in this paper may be a candidate process for
the measurement of the polarized gluon distribution function in the
proton at future polarized HERA collider.

\acknowledgments
This work was supported in part by the National Natural Science Foundation
of China, the State Education Commission of China, and the State
Commission of Science and Technology of China.

After the calculation of this paper was finished we found a paper\cite{psip},
in which a similar process has been considered, but their result is just
for polarized $\jpsi$ production at low energies.

\newpage
\appendix
\section*{}
In this appendix, we list the polarized and unpolarized cross sections
for the different partonic processes of Eq.~(\ref{cs})-(\ref{3pj}).
For convenience, we define two variables,
$P=\hat{s}\hat{t}+\hat{t}\hat{u}+\hat{s}\hat{u}$ and
$Q=(\hat{s}+\hat{t})( \hat{s}+\hat{u})( \hat{t}+\hat{u})$.

\noindent
$\gamma +g\rightarrow c\bar{c}({}^3S_1,\b{1})+g$:
the unpolarized cross section,
\begin{equation}
\frac{d\sigma}{dt}=\frac{2M(4\pi)^3\alpha\alpha_s^2e_c^2}{27\pi\hat{s}^2}
\frac{P^2-M^2\hat{s}\hat{t}\hat{u}}{Q^2}\langle {\cal O}_1^{\psi}({}^3S_1)\rangle,
\end{equation}
the polarized cross section,
\begin{equation}
\frac{d\Delta\sigma}{dt}=\frac{2M(4\pi)^3\alpha\alpha_s^2e_c^2\hat{t}\hat{u}}{27\pi\hat{s}^2}
\frac{\hat{s}^2-P}{Q^2}\langle {\cal O}_1^{\psi}({}^3S_1)\rangle
\end{equation}

\noindent
$\gamma +g\rightarrow c\bar{c}({}^3S_1,\b{8})+g$: the unpolarized and 
polarized cross sections are relevant to those for the process
$\gamma +g\rightarrow c\bar{c}({}^3S_1,\b{1})+g$ multiplied by the
factor
\begin{equation}
\frac{15}{8}\frac{\langle {\cal O}_8^{\psi}({}^3S_1)\rangle}
{\langle {\cal O}_1^{\psi}({}^3S_1)\rangle}
\end{equation}

\noindent
$\gamma +g\rightarrow c\bar{c}({}^1S_0,\b{8})+g$:
\begin{eqnarray}
\frac{d\sigma}{dt}&=&\frac{3(4\pi)^3\alpha\alpha_s^2e_c^2\hat{s}\hat{u}}
	{16\pi\hat{s}^2}
\frac{M^8+\hat{s}^4+\hat{t}^4+\hat{u}^4}{M\hat{t}Q^2}
\langle {\cal O}_8^{\psi}({}^1S_0)\rangle,\\
\frac{d\Delta\sigma}{dt}&=&\frac{3(4\pi)^3\alpha\alpha_s^2e_c^2\hat{s}\hat{u}}
	{16\pi\hat{s}^2}
\frac{M^8+\hat{s}^4-\hat{t}^4-\hat{u}^4}{M\hat{t}Q^2}
\langle {\cal O}_8^{\psi}({}^1S_0)\rangle.
\end{eqnarray}

\noindent
$\gamma +g\rightarrow c\bar{c}({}^3P_0,\b{8})+g$:
\begin{eqnarray}
\nonumber
\frac{d\sigma}{dt}&=&\frac{(4\pi)^3\alpha\alpha_s^2e_c^2}{4\pi\hat{s}^2Q^2}
\langle {\cal O}_8^{\psi}({}^3P_0)\rangle
[\frac{9M^5\hat s\hat u}{t}+
\frac{\hat s\hat u\hat t^3(2M^4+3\hat tM^2+\hat s\hat u)^2}
{M^3(\hat t+\hat s)^2(\hat t+\hat u)^2}\\
&&+\frac{\hat s^3 \hat u(3\hat s^2M^2-\hat t\hat u(2M^2-\hat s))^2}
{M^3\hat t(\hat t+\hat s)^2(\hat s+\hat u)^2}+
\frac{\hat u^3 \hat s(\hat s\hat t(2M^2-\hat u)-3\hat u^2M^2)^2}
{M^3\hat t(\hat t+\hat s)^2(\hat s+\hat u)^2}],\\
\nonumber
\frac{d\Delta\sigma}{dt}&=&\frac{(4\pi)^3\alpha\alpha_s^2e_c^2}{4\pi\hat{s}^2Q^2}
\langle {\cal O}_8^{\psi}({}^3P_0)\rangle
[\frac{9M^5\hat s\hat u}{t}-
\frac{\hat s\hat u\hat t^3(2M^4+3\hat tM^2+\hat s\hat u)^2}
{M^3(\hat t+\hat s)^2(\hat t+\hat u)^2}\\
&&+\frac{\hat s^3 \hat u(3\hat s^2M^2-\hat t\hat u(2M^2-\hat s))^2}
{M^3\hat t(\hat t+\hat s)^2(\hat s+\hat u)^2}-
\frac{\hat u^3 \hat s(3\hat u^2M^2-\hat s\hat t(2M^2-\hat u))^2}
{M^3\hat t(\hat t+\hat s)^2(\hat s+\hat u)^2}].
\end{eqnarray}

\noindent
$\gamma +g\rightarrow c\bar{c}({}^3P_1,\b{8})+g$:
\begin{eqnarray}
\nonumber
\frac{d\sigma}{dt}&=&\frac{(4\pi)^3\alpha\alpha_s^2e_c^2
\langle {\cal O}_8^{\psi}({}^3P_1)\rangle}
{2\pi\hat{s}^2Q^4M^3}
[\hat s^7(\hat t^4 + 2 \hat t^3 \hat u + 4 \hat t^2 \hat u^2 + 2 \hat t \hat u^3 + \hat u^4)
+\hat s^6(\hat t + \hat u)^2\\
\nonumber
&&(3 \hat t^3 + 7 \hat t^2 \hat u + 7 \hat t \hat u^2 - \hat u^3) 
+\hat s^5(3 \hat t^6 + 22 \hat t^5 \hat u + 60 \hat t^4 \hat u^2 + 76 \hat t^3 \hat u^3 + 36 \hat t^2 \hat u^4 \\
\nonumber
&&+ 4 \hat t \hat u^5 - \hat u^6)
+\hat s^4 (\hat t + \hat u)(\hat t^6 + 12 \hat t^5 \hat u + 46 \hat t^4 \hat u^2 + 72 \hat t^3 \hat u^3 + 32 \hat t^2 \hat u^4 + 4 \hat t 
\hat u^5 \\
\nonumber
&&+ \hat u^6)+2\hat s^3 \hat t \hat u(\hat t^6 + 10 \hat t^5 \hat u + 38 \hat t^4 \hat u^2 + 59 \hat t^3 \hat u^3 + 38 \hat t^2 \hat u^4 + 10
 \hat t \hat u^5 + \hat u^6) \\
\nonumber
&&+2 \hat s^2 \hat t^2 \hat u^2(\hat t + \hat u)(\hat t^4 + 9 \hat t^3 \hat u + 20 \hat t^2 \hat u^2 + 10 \hat t \hat u^3 + 2 \hat u^4)  
+\hat s\hat t^3 \hat u^3 (\hat t + \hat u)^2 \\
&&(2 \hat t^2 + 9 \hat t \hat u + 2 \hat u^2)
+ \hat t^4 \hat u^4(\hat t + \hat u)^3]\\
\nonumber
\frac{d\Delta\sigma}{dt}&=&\frac{(4\pi)^3\alpha\alpha_s^2e_c^2\hat t\hat u
\langle {\cal O}_8^{\psi}({}^3P_1)\rangle}
{2\pi\hat{s}^2Q^3(\hat t+\hat u)M^3}
[2\hat s^5(\hat t^2 + \hat t \hat u + \hat u^2) + 2 \hat s^4\hat t
(\hat t+\hat u)(2 \hat t+3 \hat u)\\
\nonumber
&& + \hat s^3\hat t(2 \hat t^3 + 5 \hat t^2 \hat u 
+ 2 \hat t \hat u^2 + \hat u^3)
 - 2\hat s^2\hat u(\hat t+\hat u) (2 \hat t^3 + 8\hat t^2 \hat u
+3 \hat t \hat u^2 + \hat u^3)\\
&& - \hat s\hat t\hat u(\hat t+\hat u)^2
( \hat t^2  + 7 \hat t \hat u + \hat u^2 )
 - \hat t^2 \hat u^2( \hat t+ \hat u)^3]
\end{eqnarray}

\noindent
$\gamma +g\rightarrow c\bar{c}({}^3P_2,\b{8})+g$:
\begin{eqnarray}
\nonumber
\frac{d\sigma}{dt}&=&\frac{(4\pi)^3\alpha\alpha_s^2e_c^2
\langle {\cal O}_8^{\psi}({}^3P_2)\rangle}
{10\pi\hat{s}^2Q^4M^3\hat t}
[12 s^9 u(t + u)^2 
+12 s^8 u (t + u)(5 t^2 + 7 t u + 4 u^2)
+s^7(3 t^5 \\
\nonumber
&&+ 142 t^4 u + 384 t^3 u^2 + 454 t^2 u^3 + 303 t u^4 + 96
 u^5)
+s^6(t + u)(9 t^5 + 202 t^4 u \\
\nonumber
&&+ 438 t^3 u^2 + 442 t^2 u^3 + 309 t u^4 + 
120 u^5) 
+s^5(9 t^7 + 198 t^6 u + 736 t^5 u^2 + 1200 t^4 u^3 \\
\nonumber
&&+ 1184 t^3 u^4
 + 860 t^2 u^5 + 429 t u^6 + 96 u^7)
+s^4 (t + u)(3 t^7 + 100 t^6 u + 450 t^5 u^2\\
\nonumber
&& + 720 t^4 u^3 + 688 t^3 u^4 
+ 496 t^2 u^5 + 255 t u^6 + 48 u^7)
+2 s^3u(11 t^8 + 114 t^7 u + 362 t^6 u^2\\
\nonumber
&& + 585 t^5 u^3 + 600 t^4 u
^4 + 440 t^3 u^5 + 227 t^2 u^6 + 66 t u^7 + 6 u^8) 
+2 s^2  t u^2 (t + u)^2(19 t^5 \\
\nonumber
&&+ 76 t^4 u + 104 t^3 u^2 + 84 t^2 u^3 + 48 t 
u^4 + 12 u^5)
+st^2 u^3 (t + u)^2(22 t^4 + 59 t^3 u \\
&&+ 58 t^2 u^2 + 36 t u^3 + 12 u^4)
+3 (t + u)^3 t^5 u^4],\\
\nonumber
\frac{d\Delta\sigma}{dt}&=&\frac{(4\pi)^3\alpha\alpha_s^2e_c^2\hat u
\langle {\cal O}_8^{\psi}({}^3P_2)\rangle}
{10\pi\hat{s}^2Q^3M^3\hat t(\hat t+\hat u)}
[12 s^7(t + u)^2
+12 s^6 (t + u)(4 t^2 + 5 t u + 3 u^2)
+2 s^5(41 t^4 \\
\nonumber
&&+ 99 t^3 u + 107 t^2 u^2 + 78 t u^3 + 30 u^4)
+6 s^4 (t + u)(12 t^4 + 21 t^3 u + 24 t^2 u^2 + 20 t u^3 \\
\nonumber
&&+ 10 u^4)
+s^3(30 t^6 + 117 t^5 u + 226 t^4 u^2 + 285 t^3 u^3 + 228 t^2 u^4 
+ 108 t u^5 + 24 u^6)\\
\nonumber
&&+2 s^2t (t + u)^2(2 t^4 + 16 t^3 u + 20 t^2 u^2 + 13 t u^3 + 6 u^4)
+s t^2 u (t + u)^2(7 t^3 + 5 t^2 u \\
&&- 17 t u^2 - 12 u^3)
+3 t^4 u^2 (t + u)^3].
\end{eqnarray}

In practice, the three ${}^3P_J^{(8)}$ ($J=0,~1,~2$) subprocesses 
contributions to $\jpsi$ production are always added together, by 
using the relation of their
color-octet matrix elements due to the heavy quark spin symmetry 
Eq.~(\ref{hss}).
For convenience, we also give here the total sum of the 
cross sections for the three ${}^3P_J^{(8)}$ processes, 
which is more simpler than the above expressions.

\vskip 0.4cm
\noindent
The sum of the three ${}^3P_J^{(8)}$ subprocesses,
\begin{eqnarray}
\nonumber
\frac{d\sigma({}^3P_J^{(8)})}{dt}&=&\frac{3(4\pi)^3\alpha\alpha_s^2e_c^2
\langle {\cal O}_8^{\psi}({}^3P_2)\rangle}
{2\pi\hat{s}^2Q^3M^3\hat t}
[7 \hat s^7\hat u(\hat t + \hat u) 
+\hat s^6\hat u(25 \hat t^2 + 38 \hat t \hat u + 21 \hat u^2) \\
\nonumber
&&
+\hat s^5 (\hat t + \hat u)(2 \hat t^3 + 45 \hat t^2 \hat u + 43 \hat t \hat u^2 + 35 \hat u^3)
+\hat s^4(4 \hat t^5 + 63 \hat t^4 \hat u + 132 \hat t^3 \hat u^2 \\
\nonumber
&&+ 156 \hat t^2 \hat u^3 + 98 \hat t \hat u^4 + 35 \hat u
^5)
+\hat s^3 (\hat t + \hat u)(2 \hat t^5 + 45 \hat t^4 \hat u + 91 \hat t^3 \hat u^2 + 99 \hat t^2 \hat u^3 \\
\nonumber
&&+ 57 \hat t \hat u^4 + 21 \hat u
^5)
+\hat s^2\hat u(13 \hat t^6 + 70 \hat t^5 \hat u + 136 \hat t^4 \hat u^2 + 132 \hat t^3 \hat u^3 + 88 \hat t^2 \hat u^4 \\
\nonumber
&&+ 38 \hat t \hat u^5 + 7 \hat u^6) 
+\hat s\hat t \hat u ^2(\hat t + \hat u) (13 \hat t^4 + 34 \hat t^3 \hat u + 29 \hat t^2 \hat u^2 + 18 \hat t \hat u^3 + 7 \hat u^4)
\\
&&+2 \hat t^4 \hat u^3 (\hat t + \hat u)^2],\\
\nonumber
\frac{d\Delta\sigma({}^3P_J^{(8)})}{dt}&=&\frac{-3(4\pi)^3\alpha\alpha_s^2e_c^2
\hat u
\langle {\cal O}_8^{\psi}({}^3P_2)\rangle}
{2\pi\hat{s}^2Q^3M^3\hat t}
[\hat s^7(\hat t + \hat u)+\hat s^6(7 \hat t^2 + 2 \hat t \hat u + 3 \hat u^2)+\hat s^5 (\hat t + \hat u)\\
\nonumber
&&(11 \hat t^2 - 17 \hat t \hat u + 5 \hat u^2)
+\hat s^4(9 \hat t^4 - 28 \hat t^3 \hat u - 48 \hat t^2 \hat u^2 - 26 \hat t \hat u^3 + 5 \hat u^4)\\
\nonumber
&&+\hat s^3 (\hat t + \hat u)(8 \hat t^4 - 19 \hat t^3 \hat u - 37 \hat t^2 \hat u^2 - 23 \hat t \hat u^3 + 2 \hat u^4)
+\hat s^2\hat t (2 \hat t + \hat u)\\
\nonumber
&& (2 \hat t^4 + 7 \hat t^3 \hat u - 2 \hat t^2 \hat u^2 - 19 \hat t \hat u^3 - 8 \hat u^4)
+\hat s \hat t^2 \hat u (\hat t + \hat u)(6 \hat t^3 + 6 \hat t^2 \hat u \\
&& -13 \hat t \hat u^2 - 10 \hat u^3)+2 \hat t^4 \hat u^2 (\hat t + \hat u)^2].
\end{eqnarray}

\newpage
\vskip 10mm
\centerline{\bf \large Figure Captions}
\vskip 1cm
\noindent
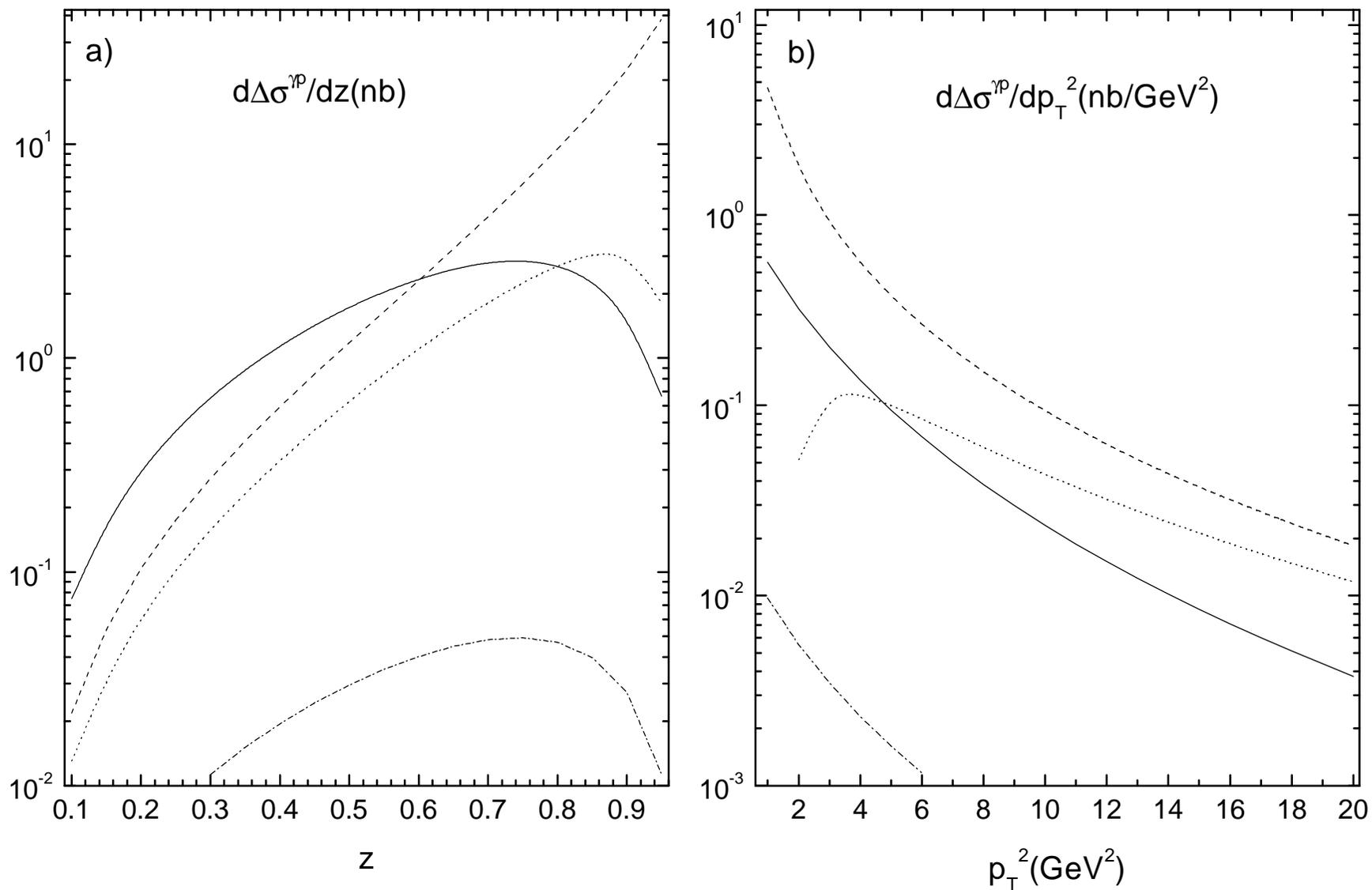
FIG. 1. The polarized cross section distribution as a function of
(a) $z$ and (b) $p_T^2$ in polarized photon-proton collisions at the
energy range of $W_{\gamma p}=100GeV$.

\noindent
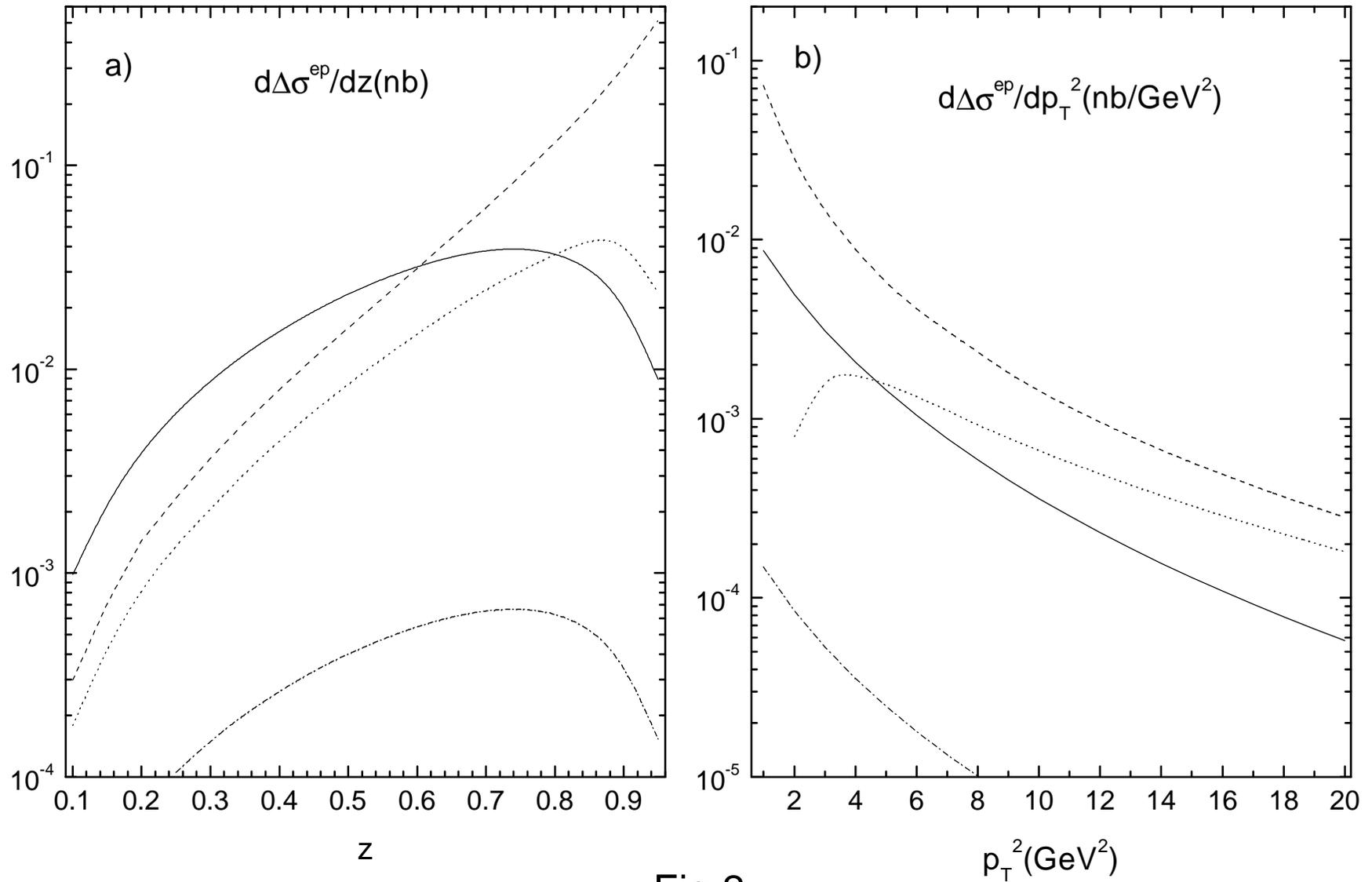
FIG. 2. The polarized cross section distribution as a function of
(a) $z$ and (b) $p_T^2$ in polarized electron-proton collisions at the
energy range relevant to HERA, $\sqrt{s}=300GeV$, where we use the
Weizs\"acker-Williams approximation to evaluate the production rate
for electron-proton collisions.

\noindent
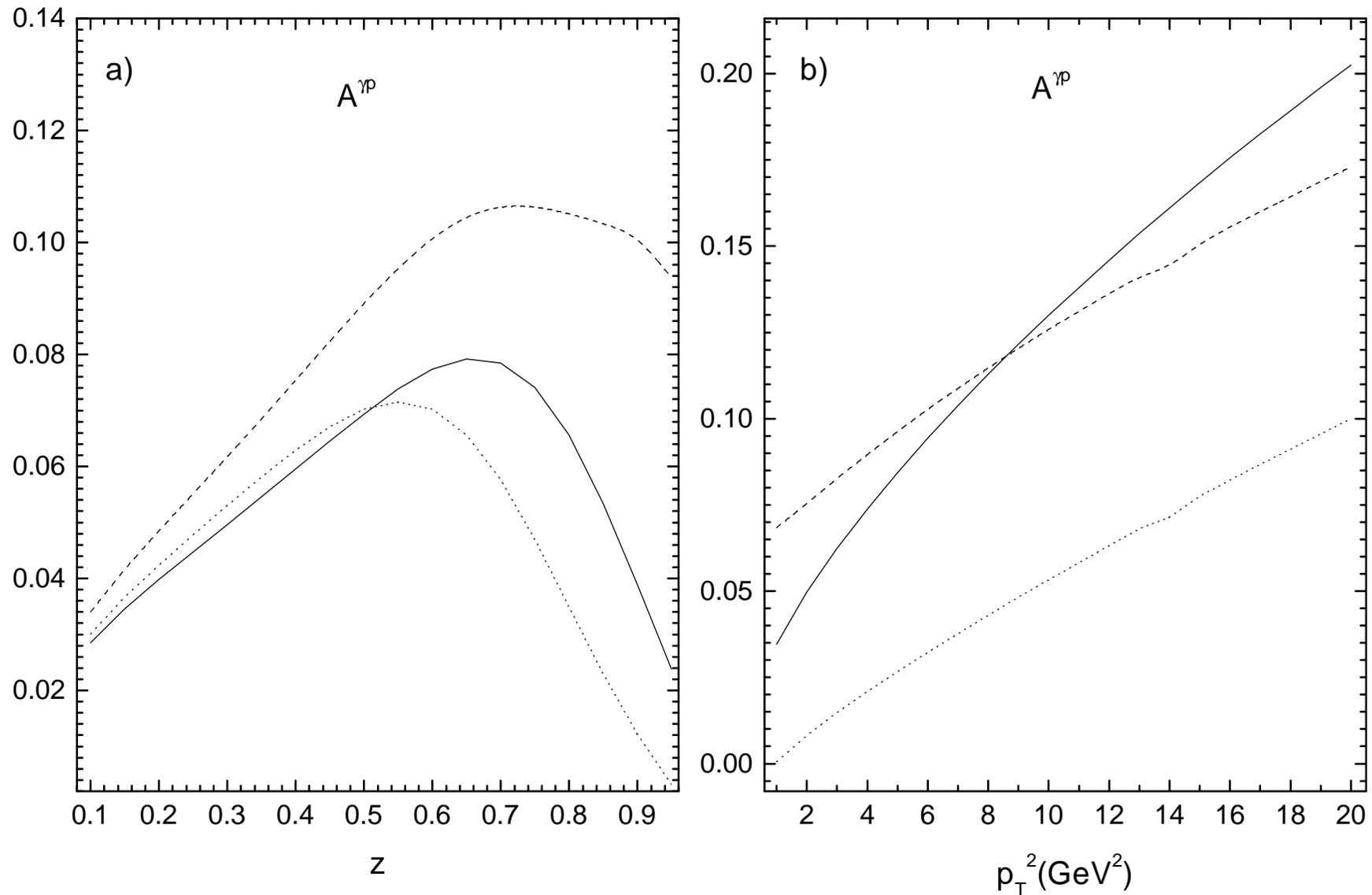
FIG. 3. The production asymmetry distributions for Fig.~1.

\noindent
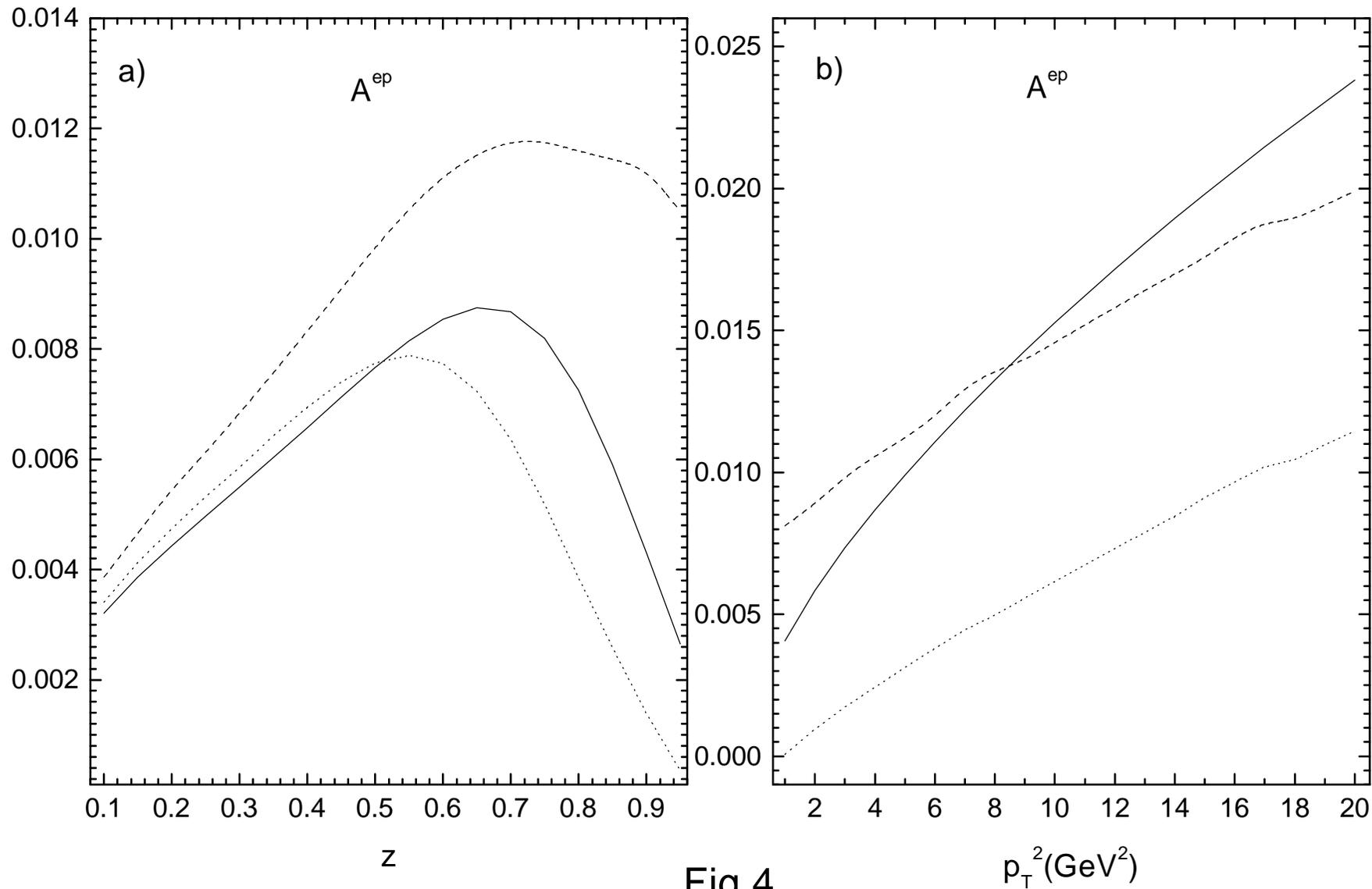
FIG. 4. The production asymmetry distributions for Fig.~2.

\noindent
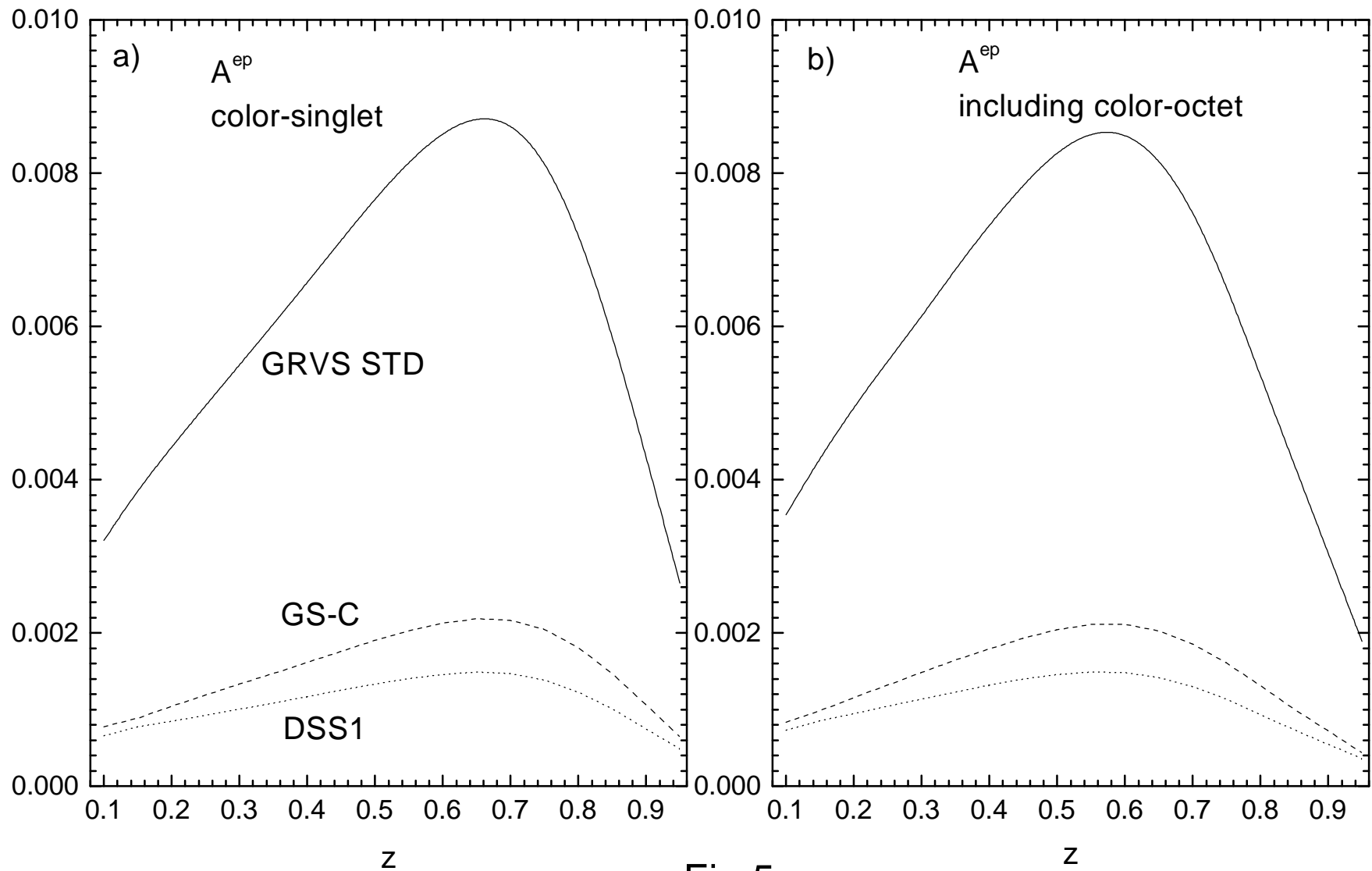
FIG. 5. The $z$ distributions of the production asymmetry $A$ for the 
inelastic $\jpsi$ photoproduction in polarized electron-proton collisions 
for different sets of polarized gluon distribution in the proton.
(a) are the results for the color-singlet predictions, and (b) are the results after including
the color-octet contributions.

\noindent
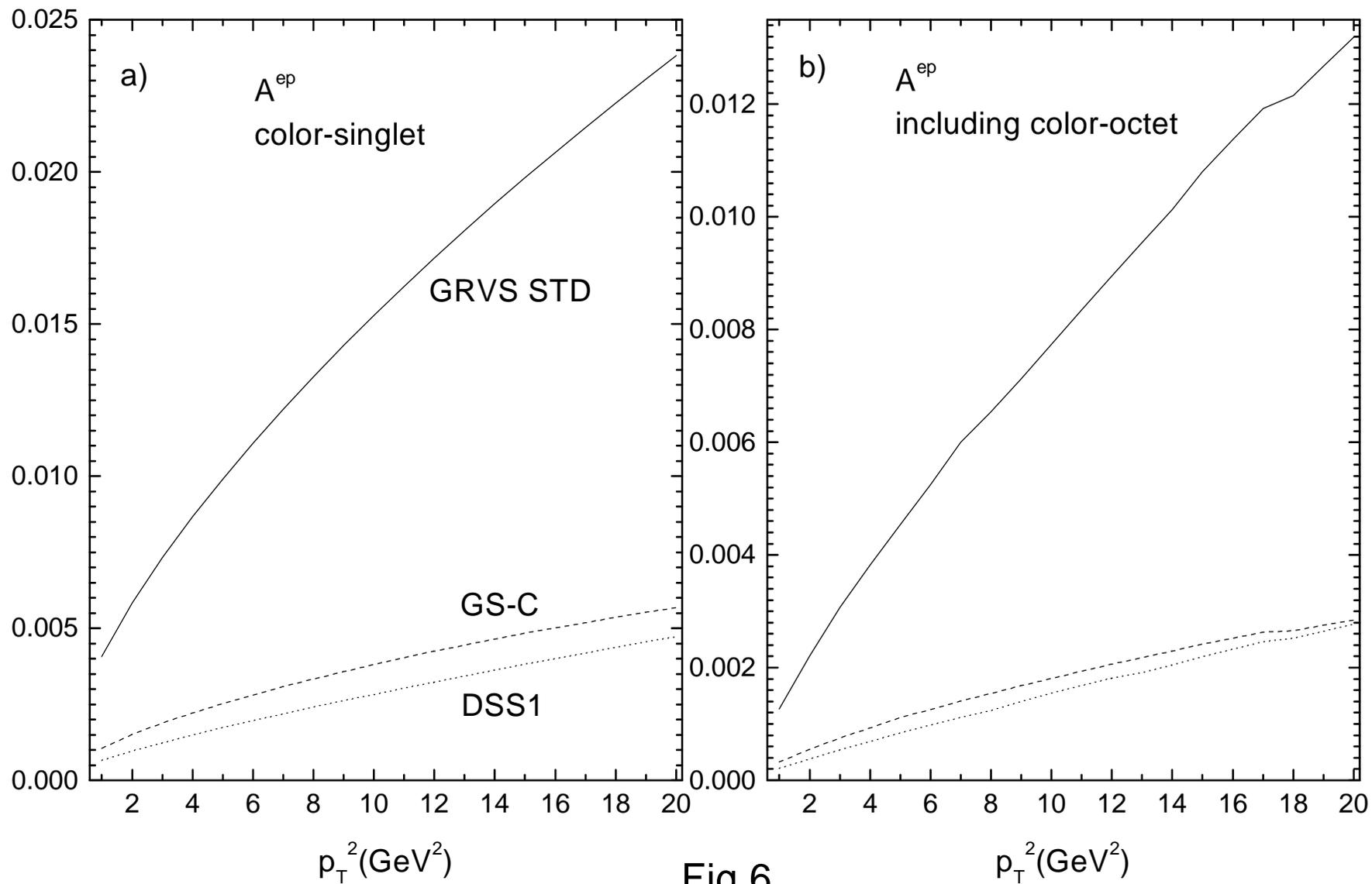
FIG. 6. The same as Fig.~5, but for the $p_T^2$ distributions.
\end{document}